# The Role of Incentives for Opening Monopoly Markets: Comparing GTE and BOC Cooperation with Local Entrants[*]

## Federico Mini[†]

## Abstract

While the 1996 Telecommunications Act requires all incumbent local telephone companies to cooperate with local entrants, section 271 of the Act provides the Bell companies—but not GTE—additional incentives to cooperate. Using an original data set, I compare the negotiations of AT&T, as a local entrant, with GTE and with the Bell companies in states where both operate. My results suggest that the differential incentives matter: The Bells accommodate entry more than does GTE, as evidenced in quicker agreements, less litigation, and more favorable prices offered for network access. Consistent with this, there is more entry into Bell territories.

---

[*] I wish to thank my dissertation advisor, Marius Schwartz, and my committee members, John Mayo, Serge Moresi and Steven Olley. I also wish to thank Scott Bohannon, Richard Clarke, Joseph Farrell, Alex Raskovich, Pierre Régibeau, Michael Riordan, and FCC seminar participants. Financial support from AT&T is gratefully acknowledged. All errors are, of course, my own.

[†] Georgetown University and Autorità per le Garanzie nelle Comunicazioni, Servizio Analisi Economiche e di Mercato, Centro Direzionale—Isola B/5, Palazzo Torre Francesco, 80143 Napoli, ITALY. *email: f.mini@agcom.it*

> *"The big difference between us and them [GTE] is*
> *they're already in long distance…what's their incentive?"*
>
> Ameritech's CEO Richard Notebaert,
> Washington Post, October 23, 1996.

## I. Introduction

The Telecommunications Act of 1996[1] aims to open all telecom markets to competition, including the large and still mostly monopolized *local* markets. Traditionally, local telephone companies in the U.S. (Incumbent Local Exchange Carriers, or ILECs) had exclusive franchise areas, in which they alone provided local exchange services and exchange access for long-distance services. Section 253 of the Act strikes down legal barriers to entry. Section 251 aims to remove artificial incumbency advantages. It requires all ILECs to provide, through an interconnection agreement, non-discriminatory access at cost-based prices to their local networks for any requesting competitor (Competitive Local Exchange Carrier, or CLEC).

In the case of the Bell Operating Companies (BOCs),[2] section 271 of the Act takes an additional step to open their local markets, beyond the obligations imposed under section 251 on all ILECs. Section 271 requires that a BOC applying for authority to offer long-distance (interLATA) services originating in a state where it offers local service (*in-region* state) must first open its local markets in the state to competition.

Section 271 and its implementation have been arguably the most contentious aspect of the 1996 Act. Supporters see 271 as an important tool for opening up local markets, because allowing a BOC into long distance before it has opened its local market would diminish its incentive to cooperate with entrants. Incentives would diminish for two reasons. Most obviously, a BOC will have less to gain from cooperating, having secured its desired entry authority (the so-called "carrot effect"). Secondly, it will have more to lose: because long-distance authority enables a BOC to offer also one-stop shopping for local and long-distance services, cooperating in opening the local market would diminish the BOC's profits in these additional markets for one-stop shopping that become available to it only after obtaining long-distance authority. Moreover, by requiring BOCs to relinquish significant control over their bottleneck facilities, section 271 reduces the risk that BOCs might use control of essential inputs to engage in anticompetitive practices in the long-distance market. Critics, however, maintain that 271 needlessly delays the BOCs' entry into long distance, because it is superfluous for opening local markets given the obligations already imposed by section 251 on all ILECs, while the BOCs pose no threat to competition in the long distance market. Less extreme critics argue that the benefits from increased local competition due to section 271 are small compared to the gains from BOC entry into long distance.[3]

Therefore, a key question in the 271 debate—and more broadly—is whether regulatory obligations, together with generic antitrust and tort law, are sufficient to induce incumbents in network industries to cooperate efficiently and expeditiously in establishing the requisite conditions for competition. If not, then incentive devices such as the 271 mechanism may play a useful role. This paper addresses that question by comparing the conduct of BOCs towards local competitors with that of GTE—by far the largest non-BOC ILEC. Unlike the BOCs, GTE has been free since the 1996 Act to offer long-distance services jointly marketed with its local services;[4] correspondingly, one expects it to be less disposed to cooperating with local entrants.

There is a widespread perception that GTE's differential incentives make it more resistant than the BOCs in dealing with local competitors. This stance is said to have included more premature litigation and tougher requests in the arbitration process, and to have resulted in less entry in GTE's territories. However, the evidence for this view has been, so far, largely anecdotal (see Burns and Kovacs [1996], [1997]).

The aim of this paper is to test the differential conduct hypothesis more systematically. Using a comprehensive, originally assembled data set, I compare the negotiations of the same local entrant, AT&T, with GTE and with the BOCs in GTE's states. AT&T is chosen because it has been one of the most active CLECs in seeking access to incumbents' networks, and has made available to me data on its negotiations. I focus on GTE because it is by far the largest non-BOC ILEC, with a size comparable to the average BOC at the time (see Mini, 2001, on AT&T's negotiation record with the other major non-BOC ILECs).

Under the Act, interconnection agreements are negotiated on a state-wide basis. My data set covers negotiations between AT&T and GTE in 23 of GTE's 28 states. In 22 of these 23 states, a BOC also offers local service (the exception is Hawaii), and my data set also covers AT&T's negotiations with the BOC in each of these 22 states. The sample therefore provides a convenient natural experiment for comparing BOC behavior with that of the largest ILEC lacking the section 271 incentives, in their negotiations with the same local entrant while facing the same state regulatory commission.[5]

Note that my analysis does not attempt to determine if more accommodation of entrants is socially superior to less accommodation. To the extent that differential incentives might have led a BOC to be more cooperative than GTE, this cooperation might be excessive, and society would have been better off had the BOC stood its ground as GTE was able to do. Expressions such as "GTE is less cooperative" are thus used only in a positive sense, and should not be construed as implying any normative assessment.

The paper is organized as follows. Section II summarizes the regulatory framework established to open up local telephone markets to competition, and describes my data set. The next two sections use originally collected data to compare GTE and BOC behavior: section III compares the negotiation process—its length and the accompanying litigation; section IV reports and analyses parties' pricing requests in arbitration, for selected terms in the interconnection contracts, and arbitrators' decisions. Section V presents the evidence about differences in entry levels in services areas served by GTE and the BOCs, based on a FCC's [1999] survey. Section VI concludes.

## II. The Interconnection Negotiation Process and the Data Set

Sections 251 and 252 of the Telecom Act require incumbent LECs to cooperate with entrants in reducing artificial barriers to local competition. Section 251 requires ILECs to enter into "interconnection agreements" with requesting CLECs to provide them: (1) interconnection to the ILEC local networks (i.e., arrangements for the exchange of traffic); (2) access to unbundled elements of ILEC networks; and (3) the ILEC's retail services at discounted wholesale rates, for resale by the entrants. Section 252 sets streamlined procedures for negotiation, compulsory arbitration absent voluntary agreement, and final approval by state public utility commissions.

Imposing such obligations on incumbents reflects a judgement that removing legal entry barriers alone would not be enough for competition to develop rapidly and efficiently in local markets (Farrell [1996]). Incumbent LECs control the local networks in their regions, and still have the vast majority of local customers, and could use this control to discourage or delay entry by limiting the cooperation they extend to entrants. The requirements imposed by the Act seek to facilitate entry through any of three modes: entirely *facilities based*; leasing from the incumbent *unbundled network elements*;[6] and through *resale* of the incumbent's existing retail services. To succeed through any of these entry modes, an entrant requires significant cooperation from the incumbent LEC.[7]

**(i) The Negotiation Process Under Section 252**

In order to speed up the process of signing such agreements, § 252 mandates a four-step negotiation procedure with highly compressed deadlines.

First, the parties may enter into a voluntary agreement within 135 days of the CLEC's request for interconnection, with any party entitled to request the state commission's mediation at any point. Second, between the 135th and the 160th day, either party may petition the state commission to arbitrate any unresolved issue (the choice of arbitration method is left to the state commission, and is discussed further in Section IV). Third, the state commission must resolve each open issue within 9 months of the original interconnection request (this has not happened in practice, as discussed shortly); the parties must then incorporate the arbitrated decision into an interconnection agreement and submit it to the state commission for final approval. Finally, the commission must approve or reject the agreement within 30 days of submission of a contract adopted through arbitration (90 days if the contract was reached by voluntary negotiation). If the commission does not act within such deadlines, the agreement is deemed approved.

No state court has jurisdiction to review a state commission's action in approving or rejecting the agreement. Any party aggrieved by the state commission's decision may bring an action in the appropriate federal court to determine if the approved agreement meets the requirements of sections 251 and 252.

Within this framework, there remain several sources for delay. First, although the Act establishes a 9-month deadline for state commissions to resolve disputed issues and produce a final arbitrated decision, the Act is silent about what happens if a state commission exceeds this deadline; the Act sets no sanctions or alternative procedures to resolve disputes in such cases. In practice, commissions have issued decisions on the disputed issues within the 9-month deadline, but the decisions were often incomplete or required parties to file additional information. In my sample, it was not at all uncommon for arbitration proceedings to be still pending nine months after the original request.

Second, even where a commission ostensibly resolves all issues, the parties must incorporate the arbitration decision into a *voluntary* contract; this leaves room for disputes over fine points. Importantly, there is no statutory deadline for the parties to submit a revised voluntary agreement after the arbitrator has resolved the disputed issues, and no provision for the commission to intervene in preparing this revised contract.

Consequently, considerably more than 9 months can elapse from the CLEC's initial request to the submission of a final contract for approval to the state commission.

**(ii) The Data Set: AT&T's Negotiations with GTE and the BOCs**

Shortly after the passage of the Act, AT&T embarked on a major national effort to enter local markets. Around March 1996, AT&T asked GTE for interconnection in 26 of the 28 states where GTE operates local networks. AT&T and GTE initially agreed to conduct negotiations at the national level, rather than for each state individually. This effort to reach agreement was unsuccessful, however, and state-by-state arbitration was invoked almost always.

My sample of negotiations between AT&T and GTE covers 23 of the 28 states where GTE offers local service.[8] In all 23 states, the negotiations ended up in arbitration. In 22 of these 23 states, a BOC also offers local service (the exception is Hawaii). I also gathered information on AT&T's negotiations with BOCs in these 22 states. Negotiations in all 22 cases ended in arbitration. Table A.1 in the Appendix lists the relevant states, the particular BOC (if any) that also provides local service in the GTE state, and the outcome of the voluntary negotiations (if any) between AT&T and that BOC. The Appendix also describes the data collection process.

Regarding contract terms, it would be impossible to account for all the price and non-price issues on which parties negotiated; the technological intricacies of local networks are mirrored by the multi-dimensional, complex nature of interconnection agreements. Thus, I focussed on a few key prices: 1) resale discount rates; 2) prices for unbundled loops, and 3) prices for end-office switching.[9] These prices are perceived as very important by the parties, at least judging from the extensive litigation record on those issues. Moreover, the quantitative nature of pricing data makes comparisons easier than for qualitative non-price issues.[10]

## III. Delay and Deadlock in Negotiations, and Premature Litigation

Section (i) below discusses the status and length of negotiations in my sample; section (ii) discusses the record on premature litigation.

Delaying an interconnection agreement with local entrants (or failing to reach an agreement), or engaging in "unjustified" litigation regarding the terms of the agreement can jeopardize a BOC's long-distance authorization by leading the state commission or the FCC to conclude that the BOC's local market is not open.  Engaging in delay or litigation to slow down entry will therefore be more costly to a BOC than to GTE, since the latter may already offer long-distance services unconditionally.[11]

### (i) Delay and Deadlock in Negotiations

*Deadlock*

As of March 1999, AT&T had obtained approved interconnections agreements with the BOCs in all my sample states except for two.  Negotiations with GTE, however, had not been finalized in 10 out of its 23 states.  Table I summarizes evidence on deadlock (Table A.2 in the Appendix provides the data by state).

**Table I—Statewide Comparison on the Status of Negotiations**

| **Number of States where:** | | | | |
|---|---|---|---|---|
| | | Approved | No Agreement | |
| **GTE has** | Approved Agreement | 12 | 0 | 12 |
| | No Agreement | 8 | 2 | 10 |
| | | 20 | 2 | 22 |

I use the McNemar statistic (see Marascuilo and McSweeney [1977]) to test the null hypothesis $H_0$: "AT&T is equally likely to get to an approved agreement with GTE and the BOC" against the alternative hypothesis $H_1$: "The likelihood of reaching an approved agreement is higher with the BOC."  The data reported in Table I leads to rejecting $H_0$ at the 1% level of significance.[12]

*Delay in Negotiations*

The length of the negotiations is measured here as the time between AT&T's request for interconnection and the approval of the final contract by the state commission.

*States where agreement was reached.* In the 12 states where AT&T has an approved agreement with both GTE and the BOC, 11 times the agreement was reached with the BOC first (see Table A.2). Using a Sign Test one can reject the null hypothesis that GTE and the BOC are equally likely to be the first to reach an approved agreement with AT&T at the 1% significance level.[13] In these 12 states, the length of negotiations between AT&T and GTE is, on average, 243 days longer than that between AT&T and the BOCs—657 and 414 days, respectively (here and below, averages are computed by assigning equal weights to each state). Using a t test for equality of means, this difference is statistically significant at the 5% level.[14]

*Entire sample.* I estimate the difference in the length of negotiations with GTE and with the BOCs *for the whole sample* by using a censored normal regression. The dependent variable vector contains *uncensored observations*—the length of the negotiations process when AT&T had reached an agreement with the incumbent LEC— and *censored observations*—where there was no approved agreement between AT&T and the ILEC as of March 1999. For the censored observations, a lower bound for the length of negotiations process was calculated as the time between AT&T's initial request and the date of the most recent statement by the parties or by the commission stating that an agreement had not yet been approved.[15] Regressing this vector on a constant and a dummy for GTE yields the following results: it takes AT&T an estimated 462 days to get an approved contract with a BOC and 853 days with GTE—eighty-five percent longer with GTE; this difference is statistically significant at the 1% level.

As compared to GTE's, BOCs' service areas in a state tend to have characteristics that make them more attractive to entrants—larger, higher subscriber density, more business customers. It has been suggested that such differences have resulted in AT&T's pursuing an interconnection agreement with the BOCs more actively than with GTE—and obtaining it more quickly. It is not evident, however, why—in the absence of the 271 incentives—the BOCs would not resist entry more than GTE if their threatened areas are more profitable, and hence why greater profitability would lead to faster agreements. Moreover, the evidence that is available is at odds with this hypothesis. First, in the overwhelming majority of my sample states, AT&T requested interconnection with GTE and the BOC within days. Second, in states like California and Texas, where the economic characteristics of GTE's and BOC's territories are reasonably similar, interconnection agreement with the BOC was reached two and six months faster, respectively.

In conclusion, while I cannot not rule out the possibility that AT&T's differential efforts might partially account for differences in the length of negotiations, there is reason to believe—especially in light of evidence (on premature litigation and pricing requests) presented later in the paper—that BOCs' stronger incentives to cooperate with entrants as compared to GTE's did play a role in the outcome and length of negotiations.[16]

**(ii)     Premature Litigation**

In order to speed up negotiations, the Act denies state courts jurisdiction to review state commissions' actions in approving or rejecting an agreement. Parties may file claims only in federal courts, and only after a state commission has issued an order approving or rejecting the arbitrated (or negotiated) interconnection agreements. "Premature litigation" is therefore used here to include claims filed prior to a final commission order, challenging either the arbitrator's decisions, or the commission's interlocutory orders.[17]

What might GTE hope to gain from such premature litigation? Premature litigation might well have caused some delay in negotiations because state commissions have limited resources, and having to defend their case in a federal court could have slowed down their other activities to open local markets pursuant to the Act. However, it is safe to assume that the *direct* delay in negotiations caused by premature litigation was not great: a censored regression where the dependent variable is the length of negotiations between AT&T and GTE indicates that there is no statistically significant difference between the cases where GTE litigated prematurely and when GTE did not.[18]

On the other hand, the *indirect* impact of GTE's premature litigation on the bargaining process might have been quite significant. While GTE likely knew that the premature claim would eventually be dismissed with little impact on the negotiation process, contesting many of the arbitrated issues signaled to AT&T that the process would be expensive, and that it should not count on the approved contract as being truly final. After all, the dismissals by the federal courts never questioned the merits of the claims (that is, whether GTE's requests were actually fair), but simply postponed such investigation to the time the action was ripe. In other words, GTE's premature claims might well have discouraged competitive entry by increasing the uncertainty facing entrants and by increasing their expected costs through signaling a tougher posture.

Table A3 in the Appendix reports premature claims filed in states where AT&T requested interconnection from both GTE and the local BOC. While GTE filed premature claims in 17 of the 23 sample states, the BOCs did so in only 3 of the 22. Of GTE's 17 premature claims, the courts dismissed 13, and GTE withdrew one.[19] All of the BOC-initiated suits were dismissed, and AT&T withdrew all of its premature motions.

Table II summarizes the information reported in Table A.3 for the 22 states where both GTE and BOC provide local service, showing that GTE litigates prematurely far more often. A McNemar test performed on this data rejects the null hypothesis that GTE and BOC are equally likely to file premature claims at the 0.1% (not 1%!) level.

**Table II—Statewide Comparison on Premature Litigation**

| Number of States where: | | | | |
|---|---|---|---|---|
| | | Yes | No | |
| **GTE litigated prematurely** | Yes | 2 | 14 | **16** |
| | No | 1 | 5 | **6** |
| | | **3** | **19** | **22** |

The disproportionate amount of GTE's premature litigation relative to the BOCs is striking, and appears to support the idea that differential incentives matter. GTE was relatively more willing than the BOCs to resort to the less predictable and slower court process as opposed to relying on private negotiation. Again, this does not imply *per se* that GTE was behaving anti-competitively.

**IV.    Parties' Pricing Requests in Arbitration and Arbitrators' Decisions**

For the selected contract terms—resale discounts, unbundled loops, and end-office switching—my data set covers not only the arbitrated awards, but also parties' initial offers. Therefore it can be used to examine whether—comparing GTE and the BOC in a given state, hence controlling for the role of state commissions—there is any systematic difference between GTE and the BOCs in their initial requests, and, if so, whether such differences are associated with different arbitration outcomes.

13### (i) Arbitration Methods

The FCC declined to establish arbitration rules, so state commissions are free to choose the arbitration mechanism to carry out their duties.

The two most common types of arbitration are *conventional arbitration* and *final-offer arbitration*. Under the former, the arbitrator is free to impose any settlement (s)he sees fit; under the latter, the parties submit final-offer to the arbitrator, who must pick one or the other. When there are two or more issues under arbitration, final-offer arbitration can take two different forms: *"package final-offer"* where the arbitrator must pick one party's final offer in its entirety, and *"issue-by-issue final-offer"*, where the arbitrator can fashion the settlement from the components of the parties' final offers.

In 18 of the 23 States in my sample, arbitration proceedings followed the conventional arbitration scheme. In the other five,[20] the state commissions ruled that the arbitrator should use "issue-by-issue" final-offer arbitration; however, if parties' final offers were clearly unreasonable or contrary to the public interest (that is, not consistent with federal and state law), the arbitrator had discretion to set contract terms different from either party's position. Thus, the arbitrators were free to revert to *conventional arbitration* as they saw fit. In four of these five states (the exception is Iowa), the arbitrators indeed reverted to conventional arbitration mechanism for at least one of the three issues I consider. In all states in my sample, therefore, the form of arbitration pursued in practice can be viewed as conventional rather than final offer.

**(ii) Parties' Pricing Requests**

Parties' pricing requests provide especially good evidence of relative aggressiveness: the terms offered for the same item by the BOC and GTE in a given state provide a direct measure of the relative stance taken in arbitration by the two types of incumbent monopolist.[21] Consistent with the differential incentive hypothesis of this paper, I find that GTE systematically makes tougher requests than do the BOCs. The evidence on arbitration awards is discussed later.

*Resale Discounts*

Resale of the incumbent LEC's services is the quickest entry mode for a competitor planning to penetrate the local market, as it requires no network investment before a subscriber base is established (in contrast to facilities-based or unbundled-elements entry).[22] Under the Act, resale discounts are supposed to reflect the retailing costs an ILEC avoids when it sells to a CLEC at wholesale rather than selling to retail customers (costs of billing, marketing, etc.). Negotiations on resale discounts were carried out separately for residential and business subscribers. Resale discounts were quoted in *percentages* off the ILEC's retail price, both in parties' requests and in arbitration awards. Since retail prices might differ between GTE and BOC territories in each state, comparing *percentage* discounts is problematic, as a given percentage discount can yield different dollar discounts depending on the prices.

To deduce the dollar discounts corresponding to the parties' requests in arbitration, and the arbitrators' awards, ideally one would apply those percentage discounts to the actual local rates (net of taxes) charged by ILECs. Unfortunately, such data are not consistently reported.

However, using data from the Hatfield Model[23] and the FCC, I computed average monthly residential and business revenues per line, for each ILEC by state.[24] I then applied the percentage discounts quoted by the parties in arbitration to these dollar revenues to estimate the implied dollar discounts. Comparing the estimated dollar discounts for GTE and the BOC in a given state without adjusting for possible differences in their avoided retailing costs is appropriate under the assumption—widely supported in the industry—that differences in retail costs between GTE and BOC territories in a given state (and probably even across states) are likely to be relatively small.

Based on these estimated dollar resale discounts, I determined which ILEC in a given state (GTE or the BOC) offered the lower resale discount to an entrant for serving residential customers, from which ILEC AT&T requested the lower resale discount, and which ILEC received a lower resale discount in arbitration. Recall that a low resale discount is desirable for the ILEC and harmful for the entrant (here AT&T). These comparisons are summarized in Table III.

Table III—Resale Discounts

|  | ILEC offering the lower discount | AT&T requested lower discount from | Arbitrator awarded the lower discount to |
|---|---|---|---|
| **Residential** | | | |
| *GTE* | *15* | *8* | *13* |
| *BOC* | *3* | *10* | *8* |
| *# of possible statewide comparisons* | *18* | *18* | *21* |
| **Business** | | | |
| *GTE* | *13* | *6* | *8* |
| *BOC* | *5* | *12* | *13* |
| *# of possible statewide comparisons* | *18* | *18* | *21* |

*Note:* Mini (2001) reports state-by-state residential and business discounts comparisons.

Given that GTE offers the lower residential discount 15 times and the BOCs 3 times, a Sign Test rejects at the 1% significance level the null hypothesis $H_0$: "GTE and the BOC are equally likely to make the tougher offer" against the alternative $H_1$: "GTE is more likely to make the tougher offer." Conversely, the record on AT&T's residential discount requests (8:10) shows that AT&T is equally likely to make the tougher request to GTE or the BOC. Similar findings arise for resale discounts to serve *business* customers (see Table III).[25]

### *Prices of Unbundled Network Elements: Loops and Switching*

The comparison of prices of unbundled network elements does not suffer from the percentage problem arising for resale discounts, since price data in arbitration were quoted directly in dollars. However, one needs to control for differences in network costs across different service areas. Especially for loops, the cost of providing service depends critically on aspects such as customer density (which affects average loop length) and territory configuration, and it can vary quite dramatically between and within states.

To do so, I use cost estimates from the Hatfield Model. The Hatfield Model provides cost estimates for loops and end-office switching for each LEC in each state. I use those company-state specific figures to compute price/cost ratios for those unbundled elements; comparisons between GTE and BOCs are then carried out employing the computed ratios (rather than the raw price data).

For my purposes, Hatfield Model's cost estimates need not be *exactly* equal to the true cost of service for any LEC in any state: as long as the Hatfield Model does not systematically over- or under-estimate BOCs' costs *relative* to GTE's, it can serve as a meaningful benchmark to control for network costs in comparing the relative aggressiveness of GTE and the BOCs in their pricing demands.

*Loops.* Table IV summarizes state-by-state comparisons between parties' requests in arbitration and arbitrators' awards regarding price/cost ratios for loops. GTE has a

17tougher position in arbitration 11 times out of 16, which allows to reject the null hypothesis of no difference relative to the BOCs, but only at the 11% significance level.[26] As with resale discounts, AT&T's requests do not seem to differ significantly between GTE and the BOCs.

Table IV—Loop Price/Cost Ratios

|  | ILEC requesting higher price/cost ratio | AT&T offered higher price/cost ratio to | Arbitrator awarded the higher price/cost ratio to |
|---|---|---|---|
| *GTE* | 11 | 6 | 10 |
|  | 5 | 8 | 11 |
| # *of possible statewide comparisons* | 16 | 14 | 21 |

*Note:* Mini (2001) reports state-by-state comparisons.

***End-office switching.*** Table V reports similar comparisons for price/cost ratios for end-office switching. Again, GTE's requests are tougher than the BOCs' (the Sign Test rejects the null hypothesis of no difference at the 10% significance level). Although AT&T may appear as "responding" by making smaller price/cost offers to GTE than to the BOCs (7 times out of 10), the null hypothesis that AT&T is equally likely to make the better offer to GTE or the BOC (against the alternative that the offer to GTE is tougher) can only be rejected at the 18% significance level.

Table V—End-Office Switching Price/Cost Ratios

| State | ILEC requesting higher price/cost ratio | AT&T offered higher price/cost ratio to | Arbitrator awarded higher price/cost ratio to |
|---|---|---|---|
| *GTE* | 6 | 3 | 6 |
|  | 1 | 7 | 10 |
| # *of possible statewide comparisons* | 7 | 10 | 16 |

*Note:* Mini (2001) reports state-by-state comparisons.

Finally, I pooled the data on loops and switching and carried out Sign Tests on this pooled sample of unbundled network elements. The 17 times out of 23 where GTE rather than the BOC requested higher price/cost ratios allow rejection the hypothesis of equal conduct at the 5% significance level. AT&T made "tougher" (lower) requests to GTE 15 times out of 24, which is not enough to reject the hypothesis that AT&T treated BOCs and GTE equally at the 10% significance level.

Note that the evidence on parties' pricing requests in shows that AT&T's stance did not significantly differ between its negotiations with BOCs and GTE. Thus, the data in my sample do not support the suggestion that A&T longer delay in reaching agreements with GTE or its failure to do so (Section III(i)) is due to differential conduct by AT&T toward the BOCs versus toward GTE.[27]

### (iii)   Arbitration Awards

The data in Tables III to V can be used to test if GTE's tougher requests result in correspondingly more favorable arbitration awards to it than to the BOC. Interestingly, they do not. Combining business and residential resale discounts (Table III), GTE received a superior award in exactly half of the cases. Moreover, applying the Sign Test to residential and business resale discounts separately, the hypothesis of equal treatment by the arbitrator cannot be rejected at the 10% significance level. Similarly, there is no systematic difference in GTE and BOC awards for loops and switches, both when these data are analyzed separately (Tables IV and V) and when they are pooled.

The fact that GTE's awards were no better than the BOCs' might lead to conclude that its tougher requests were inconsequential. Such a conclusion, however, would overlook a potentially important effect: a tougher request by GTE could result in better awards to *both* it and the BOC in the state. In all my sample states, when a commission determined arbitration awards, it had access to both the GTE and BOC requests. Under the assumption that a commission might be reluctant to treat large ILECs—GTE and the BOC—very differently, one would not expect GTE's requests to yield significantly better awards than those received by the BOC. Nevertheless, tougher requests by GTE could induce better awards to both it and the BOC in that state. Focusing only on the difference would not capture this effect. To test this hypothesis, I examined how a request by one ILEC (GTE or the BOC) affects both its award and the award to the other ILEC.

*Resale discounts.* This is the contract term for which I have the most data points. I regressed the arbitrated dollar discount awarded to the BOC in each state (ARBBOC) on a constant, AT&T's requests to the BOCs (ATTBOC), and *both* the BOC's and GTE's requests in that state (BOC and GTE respectively). Table VI reports Ordinary Least Squares results for this regression. All variables are expressed in logarithms, so the estimated coefficients represent elasticities.

**Table VI—OLS Regressions Explaining Arbitrated Awards: Resale Discounts**

| Dependent Variable: ARBBOC | | Dependent Variable: ARBGTE | |
|---|---|---|---|
| Constant | 0.326* | Constant | 0.283 |
|  | (2.891) |  | (1.175) |
| BOC | 0.286* | GTE | 0.480* |
|  | (4.371) |  | (3.094) |
| ATTBOC | 0.438* | ATTGTE | 0.495** |
|  | (5.444) |  | (2.685) |
| GTE | 0.189* | BOC | 0.044 |
|  | (2.859) |  | (0.532) |
| R-squared | 0.927 | R-squared | 0.926 |
| Adjusted R-squared | 0.919 | Adjusted R-squared | 0.918 |
| No. of Observations | 33 | No. of Observations | 33 |

Note: t-tests in parentheses; * significant at the 1% confidence level; ** significant at the 5% confidence level.

In the regression explaining the awards to the BOCs, the coefficients on BOC and ATTBOC have the expected sign (the award is positively related to both parties' positions), and are strongly significant. The coefficient on GTE's request is also positive and strongly significant, and has the same order of magnitude as the coefficients on AT&T's and BOC's variables. (AT&T's behavior in the negotiations with GTE, instead, does not help explain the award in the BOC proceedings. In fact, when AT&T's resale requests to GTE were included in the regression—either together with GTE's requests or by themselves—the associated coefficient was not significantly different from zero at the 5% confidence level.) Note also that, notwithstanding the small size of the sample, the regression explains almost 93% of the variation in the data.

The regression explaining the resale discount awarded to GTE (ARBGTE) reveals an interesting contrast. While the coefficients on AT&T's and GTE's requests have the expected sign and are strongly significant, that on BOC's request is statistically insignificant. (Again, when AT&T's resale requests to the BOC in the state were included as an extra explanatory variable for the GTE—either together with BOC's requests or by themselves— the associated coefficient was not significantly different from zero at the 10% confidence level.) I offer a possible explanation for the above results shortly.

*Loop price/cost ratios.* Table VII reports similar regressions explaining BOC and GTE arbitrated awards for loop price/cost ratios. Again, all variables are expressed in logarithms. (The same variable names used above now identify logarithms of price/cost ratios). Regarding BOC awards, while the coefficients on BOC and ATTBOC are not statistically significant, GTE's requests appear to affect the arbitrator's award to the BOCs in the same way found above for resale discounts. (Similarly to the case of arbitrated resale discounts, ATTGTE has no explanatory power—either when included together with GTE or when GTE is excluded from the regression.)

**Table VII—OLS Regressions Explaining Arbitrated Awards: Loop Price/Cost Ratios**

| Dependent Variable: ARBBOC | | Dependent Variable: ARBGTE | |
|---|---|---|---|
| Constant | 0.060 | Constant | -0.339 |
| | (0.539) | | (-1.155) |
| BOC | -0.143 | GTE | 0.603** |
| | (-1.361) | | (2.045) |
| ATTBOC | 0.307 | ATTGTE | 0.445** |
| | (1.286) | | (1.920) |
| GTE | 0.288* | BOC | 0.130 |
| | (2.421) | | (0.624) |
| R-squared | 0.566 | R-squared | 0.685 |
| Adjusted R-squared | 0.421 | Adjusted R-squared | 0.590 |
| No. of Observations | 13 | No. of Observations | 14 |

*Note*: t-tests in parentheses; * significant at the 5% confidence level; ** significant at the 10% significance level.

In the regression explaining GTE awards, the BOC's request is clearly insignificant, while GTE's and AT&T's are significant at the 10% level. (Again, the coefficient on ATTBOC is not statistically significant, either when included together with BOC or when it replaces this latter regressor.)

A possible interpretation of these findings is as follows. The state commission pays more attention to GTE's request, recognizing that the BOC is less willing to press its case (through legal challenges and otherwise) because the BOC is concerned that such challenges might adversely affect its prospect for securing long-distance entry authority.[28] The reluctance to award significantly different prices to GTE as compared to the BOC in the same state, however, results in both ILECs receiving higher awards than they would have had GTE made a "softer" request.

Could it be the case that, in determining the arbitrated award to GTE and the BOC respectively, state commissions pay attention *only* to GTE requests? This interpretation is consistent with the results reported in Table VII—where AT&T's and the BOC's position do not explain the arbitrated loop price/cost ratio in the BOC proceedings. On the other hand, the fact that in the regression explaining resale discounts to the BOC (Table VI) the coefficients on the variables BOC and ATTBOC are statistically significant might simply pick up differences between BOC and GTE markets.

To address this question, I re-estimated the regressions explaining the arbitrated resale discounts in the BOC proceedings reported in Table VI and VII including number of lines and subscribers' density in BOC service areas as proxies for differences in market characteristics. The results[29] show that, once market characteristics are accounted for, the arbitrated resale discount in the BOC proceedings still depends on AT&T's and the BOC's requests *and* GTE's position in that same state. In conclusion, while further work on an explicit model of the interaction among entrants, ILECs with and without long-distance authority and arbitrators might yield interesting insights, there appears to be evidence that GTE's relatively "tougher" bargaining position was not inconsequential.

## V. Evidence on Competitive Entry Into Local Telephone Markets

GTE's less cooperative stance (more premature litigation, longer negotiations, and tougher pricing requests) is likely to discourage entry in its territories more than into BOC territories. Beyond the obstacles documented here (principally, delays or failure to reach interconnection agreements), GTE's less cooperative stance is also likely to include non-price conduct not explored in this paper but widely viewed as important (inferior technical platform arrangements, constraints on bundling of networks elements, etc.). This section presents some evidence that GTE in fact has experienced less competitive entry, after controlling for other economic factors that might influence entry decisions.

In early 1998, the FCC's Common Carrier Bureau started collecting evidence on entry into local telephone markets, by asking ILECs and CLECs to provide—on a voluntary basis—data on various competition-related issues on a state-by-state basis. The first "Survey on the State of Local Competition" was published in February 1998; it reported information as of December 31, 1997, voluntarily provided by five BOCs (Ameritech, Bell Atlantic, Bell South, Southwestern Bell, and US West), GTE, and three other large LECs (SNET, Frontier, and Sprint).[30] Since February 1998, the FCC has been revising and updating this survey periodically. In August of 1999, the FCC released an updated survey reporting revised information as of December 31, 1998.[31]

It should be noted that my arbitration data refers to *a single entrant*, AT&T, whereas the FCC's information on entry pertains to *all* entrants. The evidence that follows is thus offered only as suggestive of the difference in the ease of entry into GTE versus BOC territories, which may or may not be due to the same difficulties AT&T experienced (see sections III and IV).

**(i)    Competitive Entry Record**

*Resale Discounts*

Each carrier reported the number of switched access lines (broken down in: (a) residential, or (b) business and other lines) sold to competing carriers for resale, as a percentage of the carriers' switched lines in the state. The information as of December 1998 for BOCs and GTE in my sample is reported in Table A4 in the Appendix.

I test the null hypothesis $H_0$ against the alternative $H_1$:

$H_0$: "BOC and GTE are equally likely to have the larger % of resold lines in the state."

$H_1$: "BOC is more likely than GTE to have larger % of resold lines in the state."

The same hypotheses are tested subsequently replacing "*% of resold lines*" by "*% of unbundled loops*" (Table A5) and by "*% of ILEC lines served by switching centers where competitors have collocation*" (Table A6).

For *residential lines*, a Sign Test rejects $H_0$ against $H_1$ at the 1% significance level.[32] Applying the Wilcoxon Test (see Hogg and Craig [1978]), which takes into account the magnitude of differences of pair-wise comparisons between BOC and GTE, $H_0$ can be rejected at the 1%.[33]

For *business and other lines*, $H_0$ is rejected without the need for any formal test since the BOC has the greater % in all cases.

*Unbundled Network Elements*

For *unbundled loops*, the Sign Test rejects $H_0$ at the 1% significance level. Moreover, using the Wilcoxon test, $H_0$ can be rejected at the 5% significance level.[34] As for *collocation arrangements* between ILECs and new entrants, $H_0$ is rejected without the need for any formal test since the BOC has the greater % in all cases.[35]

**(ii)   Explaining Differences in Competitive Entry**

Part of the observed differences in entry patterns could be due to factors other than GTE's weaker incentives to cooperate with entrants. ILECs often claim that entrants are mainly drawn by the prospect of "cream-skimming" customers to whom prices are well above cost (especially business customers), and stay away from areas where subscribers' density is low (and the cost of developing own networks is higher). Areas with relatively high per-line expenditure on toll and other non-local telephone services also could be more attractive to entrants, as the revenues from provision of one-stop shopping and integrated services are likely to be larger there. Moreover, there might be large fixed entry costs (e.g., fixed set-up costs, advertising), so it would not be surprising to see more entry into relatively larger ILECs' territories—such as those of the BOCs'.

The lesser entry into GTE's region in a state as compared to the BOC's documented above could thus be due to GTE's territories, on average, being relatively (a) smaller, (b) more rural, (c) with subscribers spending less on telecommunications, and (d) with a local tariff structure closer to costs which makes it a less attractive competitive target. I investigate this issue using the FCC survey data as of December 1998, where the unit of observation is an incumbent LEC in a state. There were 79 such observations: 48 for the BOCs, the other 31 accounted for by GTE (15), Sprint (15), and SNET (1).[36] I use the 31 observations on non-BOC ILECs as the control group to investigate to what extent the differential cooperation incentives for them and the BOCs might account for the differences in competitive entry.[37]

Using the regressors in table VIII, I estimate distinct regressions explaining resale and collocation penetration in the residential and business market respectively.

**Table VIII—Explanatory Variables in Entry Regressions**

| Variable | Definition and Source |
|---|---|
| RESLINES | Log of total residential lines ('000s), Hatfield Model |
| BUSLINES | Log of total business switched lines ('000s), Hatfield Model |
| RESLOWDENS | Log of % of residential lines in zones with less than 850 line per square mile, Hatfield model |
| BUSLOWDENS | Log of % of business lines in zones with less than 850 line per square mile, Hatfield model |
| RESMARKUP | Log of residential tariff/cost ratio. |
| BUSMARKUP | Log of residential tariff/cost ratio. |
| TOLL | Log of originating toll calls ('000s) per line/month (intra and interLATA), Hatfield Model. |
| BOC | Dummy taking value 1 if company is BOC, 0 otherwise |

Note that for all explanatory variables with the exception of TOLL, data are available separately for residential and business lines for each ILEC in a state. Unfortunately, data on telecommunications expenditure per-line broken down by state, ILEC (e.g., GTE and BOC), and type of customer (residential vs. business) are not directly available. Data are available on the number of originating toll calls (TOLL) by state and ILEC—but not the breakdown of such figure in residential and business calls. I use TOLL as a proxy for the amount of per-line telephone expenditures on non-local services in both the regressions explaining residential and business entry.[38]

I also experimented with including explanatory variables at the state level—such as average median income in the state, percentage of state population in given age groups, and percentage of people above a certain education level.[39] In contrast with other regressors, which vary *in any given state* from an ILEC to the other, for these explanatory variables, the same value applies to different ILECs in the same state (e.g., GTE and BellSouth in Florida). These explanatory variables consistently turned out to be statistically insignificant, so I omitted them.

*Entry Trough Resale*

Table IX reports the results obtained regressing the variables on Table VIII on the log of the number of residential (business) resold lines per thousands of total residential (business) lines as reported by the FCC (RESRESOLD and BUSRESOLD respectively).[40]

**Table IX—OLS Regressions Explaining Entry Through Resale**

| Dependent Variable: RESRESOLD | | Dependent Variable: BUSRESOLD | |
|---|---|---|---|
| Constant | -8.826** | Constant | -12.363* |
|  | (-2.614) |  | (-5.402) |
| RESLINE | 1.751* | BUSLINE | 1.536* |
|  | (5.909) |  | (7.123) |
| RESLOWDENS | -0.310 | BUSLOWDENS | 0.315 |
|  | (-1.009) |  | (0.781) |
| RESMARKUP | -0.8102 | BUSMARKUP | 1.853*** |
|  | (-0.635) |  | (1.737) |
| TOLL | 0.601 | TOLL | 0.583 |
|  | (-1.225) |  | (1.203) |
| BOC | -0.080 | BOC | 1.487* |
|  | (-0.122) |  | (2.786) |
| R-squared | 0.5023 | R-squared | 0.7061 |
| Adjusted R-squared | 0.4640 | Adjusted R-squared | 0.6845 |
| No. of Observations | 71 | No. of Observations | 74 |

*Note:* t-tests based on robust standard errors in parentheses; * significant at the 1% confidence level; ** significant at the 5% confidence level; *** significant at the 10% confidence level.

In the residential segment, there appears to be no difference between BOCs and non-BOC ILECs when other economic factors potentially affecting entry are taken into account. The only significant coefficient is the one for ILECs' size, which is positive and significantly greater than one. This indicates that entry through resale has targeted larger service areas. In interpreting the results that all other factors turned out to be insignificant, one needs to take into account that, as of December 1998, reporting ILECs resold to competitors only 1% of their switched residential lines.

By contrast, there were three times as many resold business lines as resold residential ones, and the results for business resold lines show that entry in BOC areas is statistically greater than that into areas served by non-BOC ILECs. This gives some support to the notion that the difference in entry records cannot be fully explained by differences in economic characteristics between BOCs and non-BOCs, and that differential incentives induced by section 271 are indeed playing a role. Regarding the other regressors, ILECs' size is still positively related to entry (with a coefficient statistically greater than one). In addition, price-cost margins—larger tariff/cost ratios—appear to be positively correlated with entry. As was found for residential resold lines, subscribers' density and the proxy for per-line toll expenditure are not significantly correlated with entry.

Finally, note that both regression were based on fewer than 79 observations. This is because some non-BOC ILECs reported zero entry in their regions.[41] The fact that the samples are non-randomly selected might result in a bias in the fitted parameters in the regression, as they confound the behavioral parameters of interest with those in the relationship determining the probability of positive entry.

To account for this issue, I used the Heckman two-step procedure (Heckman [1979]) to estimate the regressions above in a two-equation framework, where a probit equation explaining entry is employed to correct the sample selection bias. For all specifications of the probit regression explaining entry, I could not reject the hypothesis of no sample selection bias, based on the Lagrange multiplier test statistic proposed by Melino (1982). Thus, no sample selection bias correction was applied.[42]

*Entry Through Unbundled Network Elements*

Table X reports the regression results where the extent of entry through UNE is proxied by the log of the percentage of residential (business) lines in switching centers where competitors have collocation arrangements (PCRESCOLL and PCBUSCOLL, respectively).[43]

**Table X—OLS Regressions Explaining Entry Through UNE**

| Dependent Variable: PCRESCOLL | | Dependent Variable: PCBUSCOLL | |
|---|---|---|---|
| Constant | -3.651 | Constant | 0.239 |
|  | (-1.543) |  | (0.238) |
| RESLINE | -0.002 | BUSLINE | 0.105*** |
|  | (-0.016) |  | (1.670) |
| RESLOWDENS | -0.512*** | BUSLOWDENS | -0.563** |
|  | (-1.926) |  | (-2.561) |
| RESMARKUP | 1.094** | BUSMARKUP | 0.946** |
|  | (2.483) |  | (2.242) |
| TOLL | 0.154 | TOLL | 0.159 |
|  | (0.834) |  | (0.957) |
| BOC | 0.887* | BOC | 0.476** |
|  | (3.450) |  | (2.415) |
| R-squared | 0.5944 | R-squared | 0.6176 |
| Adjusted R-squared | 0.5617 | Adjusted R-squared | 0.5868 |
| No. of Observations | 68 | No. of Observations | 68 |

*Note:* t-tests based on robust standard errors in parentheses; * significant at the 1% confidence level, ** significant at the 5% confidence level; *** significant at the 10% confidence level.

Results are broadly consistent across the residential and business regressions. Higher price-cost margins are associated with higher entry levels for both residential and business subscribers, while lower density is significantly correlated with less entry for both classes of subscribers. ILEC size plays no significant role in explaining residential entry; for the business segment, entry levels appear to increase with size, but at a decreasing rate. Finally, the proxy for toll expenditure is not correlated with either residential and business entry.[44] Most important for this paper, the BOC dummy is positive and strongly significant in both regressions: controlling for other factors, there is more entry in BOCs' service areas.

The empirical evidence on entry thus seems to support the hypothesis that the BOCs' stronger incentives to cooperate with local entrants do matter for entry modes that are most dependent on ILECs' cooperation—resale and UNEs as opposed to entirely facilities-based entry.

**VI. Conclusions**

The findings reported in the previous sections support the hypothesis that GTE's weaker incentives to cooperate with local entrants—arising because GTE, but not the BOCs, already may offer long-distance services unconditionally—have a significant impact on its conduct towards entrants.

In particular: 1) GTE engages in significantly more premature litigation than do the BOCs; 2) It takes significantly longer to obtain an interconnection agreement with GTE than with the BOCs; 3) GTE's pricing requests are consistently "tougher;" 4) Tougher GTE requests are not associated with systematically better awards to GTE than to the BOC; however, a tougher GTE request is associated with better awards to both it and the BOC in that state (hence a worse outcome for the entrants); 5) An FCC (1999) survey shows less competitive entry into GTE than into BOC territories. After controlling for standard economic variables likely to influence profitability of entry, econometric analysis shows statistically significant higher levels of entry in BOC areas as compared to non-BOC ILECs', consistent with the differential incentives hypothesis.

The fact that GTE experiences less entry can reasonably be attributed to its greater resistance to cooperate with entrants. Beyond factors documented here (points 1-3 above), this resistance is also likely to involve non-price variables that are not explored in this paper but that are important to entrants.

GTE's greater resistance arises despite the fact that all incumbent LECs are required, under section 251 of the Telecom Act, to cooperate fully in opening up their local markets to entry. Unless the difference can be attributed entirely to inherently greater aggressiveness by GTE—of which there is little supporting evidence prior to the 1996 Act—a reasonable inference is that section 271 of the Act provides at least part of the explanation. Section 271 creates differential incentives by requiring the BOCs, but not GTE, to cooperate in opening their local markets to competition as a pre-condition for themselves being permitted to offer long-distance services originating in their regions.

My findings therefore suggest that regulatory sticks alone are not sufficient to encourage rapid and extensive cooperation by incumbents to open up their markets, and that additional incentive devices, such as those provided by section 271, can have a meaningful impact on incumbents' conduct and therefore on prospects for competitive entry. However, it bears reiterating that these findings, by themselves, cannot establish that section 271 is socially beneficial. To address this question would require examining issues such as whether BOC cooperation is "excessive" (e.g., below-cost prices for network access) and the costs (if any) due to delaying BOC entry into long-distance. These issues are beyond the scope of this paper, but warrant further research.

**Appendix**

GTE provides local service in 28 states; my sample covers 23 of them. Table A1 lists all GTE states, the BOC in each state, and the outcome of the voluntary negotiations involving AT&T as a local entrant.

Assembling data on interconnection negotiations presented several difficulties. First, it involved collecting information from a large number of different sources: state commissions, parties' state coordinators, associations of industry members, and federal bodies.

Second, the vast majority of the information comes from lengthy legal documents, including: parties' memos and exhibits presented during arbitration; arbitrators' decisions; commission orders (ratifying, or modifying some of the arbitrators' decisions); and parties' complaints about arbitrators' or commissions' decisions. These documents had to be searched for relevant dates, parties' requests, and final decisions about contract terms; and this information then had to be recorded in an electronic format. The rest of the information was gathered by phone or via e-mail from the parties involved; or from the internet, especially from state commissions' web pages. The National Association of Regulatory Utility Commissioners keeps a list of them (www.naruc.org/stateweb.htm).

The data collection was hindered by the fact that, in general, information about negotiations is proprietary. This is true both of the parties' requests going into arbitration, and of the terms in the arbitrated interconnection agreement. In many instances, litigants were able to obtain from state commissions the right not to make the terms in the final contracts known to third parties, even though the contracts are the object of state commissions' orders, which are themselves public documents. As a result, assembling a totally complete data set proved to be impossible.

Even when information was available, in a few cases some inconsistencies emerged, which required additional investigation. For instance, dates for the same order from different sources did not always match, or the incumbent LEC reported arbitrated prices different from AT&T's understanding of the arbitrator's decision. Therefore, where possible, each piece of information was cross-checked for reliability.

Finally, after collecting the raw data, additional work was required to enable meaningful comparison between pieces of information reported in different levels of aggregation. Take, for instance, the case of loop prices. In some cases a state-wide average loop price is reported; in others, no state-wide figure is available, and loop prices are quoted at different levels of disaggregation (usually by density zones and/or by type of physical support, e.g., 2-wire or 4-wire).

Price/cost ratios for loops are computed based on average figures that capture the typical ILEC price (cost) for a loop across all loop-density areas and quality of service (analog or digital). When the parties involved (ILECs, AT&T, and the arbitrators) failed to quote directly an average figure, I derived it as a weighted average of the prices (by density areas and/or quality of services) using the breakdown of lines across the relevant classification as reported by the Hatfield Model.

As for end-office switching price/cost ratios, I computed them as follows. The Hatfield Model reports for each ILEC in state the total number of end-office switching actual minutes and the corresponding total cost of providing such service. I used these figures to compute an average per-minute cost. End-office switching prices were usually quoted as a two-part tariff (a port fee plus a per-minute charge). I used the total number of switched lines and total number of end-office switching minutes reported in the Hatfield model to derive the typical per-minute revenue as implied by the two-part tariff.

**Table A1. GTE States and Negotiations Outcomes**

| GTE State | Negotiations AT&T – GTE | BOC in the State | Negotiations AT&T – BOC |
|---|---|---|---|
| Alabama | Arbitration | Bell South | Arbitration |
| Alaska* | No request for interconnection [A] | No BOC | Not Applicable |
| Arizona* | Adopt California proceedings [B] | US West | Arbitration |
| Arkansas* | See note [C] | Southwestern Bell | Arbitration |
| California | Arbitration | Pacific Telesis | Arbitration |
| Florida | Arbitration | Bell South | Arbitration |
| Hawaii | Arbitration | no BOC | Not Applicable |
| Idaho* | See note [D] | US West | Arbitration |
| Illinois | Arbitration | Ameritech | Arbitration |
| Indiana | Arbitration | Ameritech | Arbitration |
| Iowa | Arbitration | US West | Arbitration |
| Kentucky | Arbitration | Bell South | Arbitration |
| Michigan | Arbitration | Ameritech | Arbitration |
| Minnesota | Arbitration | US West | Arbitration |
| Missouri | Arbitration | Southwestern Bell | Arbitration |
| Nebraska | Arbitration | US West | Arbitration |
| Nevada* | No request for interconnection [E] | Pacific Telesis | Arbitration |
| New Mexico | Arbitration | US West | Arbitration |
| N. Carolina | Arbitration | Bell South | Arbitration |
| Ohio | Arbitration | Ameritech | Arbitration |
| Oklahoma | Arbitration | Southwestern Bell | Arbitration |
| Oregon | Arbitration | US West | Arbitration |
| Pennsylvania | Arbitration | Bell Atlantic | Arbitration |
| S. Carolina | Arbitration | Bell South | Arbitration |
| Texas | Arbitration | Southwestern Bell | Arbitration |
| Virginia | Arbitration | Bell Atlantic | Arbitration |
| Washington | Arbitration | US West | Arbitration |
| Wisconsin | Arbitration | Ameritech | Arbitration |

*Note:* * GTE states not included in the sample.

[A] Due to GTE's limited presence in Alaska, AT&T did not ask for an interconnection agreement. Source: Robert Mahini, Sidley & Austin, Attorneys for AT&T. E-mail dated Feb. 16, 1998 (rmahini@sidley.com).

[B] The parties have stipulated they will accept the Arbitration Decision in California, but the contract has not been finalized. Source: Robert Mahini, Sidley & Austin. E-mail dated Feb. 16, 1998.

[C] AT&T petitioned for arbitration, but then requested withdrawal of its petition in April 1997. The Petition was granted on Apr. 9, 1997. Source: Jim Moore, AT&T Law Division, fax dated Apr. 30, 1998.

[D] AT&T withdrew from arbitration, and has no plan to resume negotiations. Source: Robert Mahini, Sidley & Austin. E-mail dated Feb. 16, 1998.

[E] AT&T does not plan to ask for interconnection in the foreseeable future. Source: Robert Mahini, Sidley & Austin. E-mail dated Feb. 16, 1998.

36**Table A2. Negotiation Data**

| State | Approved Interconnection Agreement by Mar. 1999? | | AT&T Reached Agreement Faster with BOC or GTE? | BOC in the State |
|---|---|---|---|---|
| | BOC | GTE | | |
| Alabama | Yes | No | BOC* | BellSouth |
| California | Yes | Yes | BOC | PacBell |
| Florida | Yes | Yes | BOC | BellSouth |
| Hawaii | Not applicable | Yes | Not applicable | Not applicable |
| Illinois | Yes | No | BOC* | Ameritech |
| Indiana | Yes | No | BOC* | Ameritech |
| Iowa | Yes | Yes | BOC | US West |
| Kentucky | Yes | No | BOC* | BellSouth |
| Michigan | Yes | No | BOC* | Ameritech |
| Minnesota | Yes | Yes | BOC | US West |
| Missouri | Yes | Yes | BOC | SBC |
| Nebraska | Yes | Yes | GTE | US West |
| New Mexico | No | No | Not applicable | US West |
| North Carolina | Yes | Yes | BOC | US West |
| Ohio | Yes | Yes | BOC | Ameritech |
| Oklahoma | Yes | No | BOC* | SBC |
| Oregon | Yes | Yes | BOC | US West |
| Pennsylvania | No | No | Not applicable | Bell Atlantic |
| South Carolina | Yes | No | BOC* | BellSouth |
| Texas | Yes | Yes | BOC | SBC |
| Virginia | Yes | No | BOC* | Bell Atlantic |
| Washington | Yes | Yes | BOC | US West |
| Wisconsin | Yes | Yes | BOC | Ameritech |
| Totals | 20 out of 22 | 13 out of 23 | | |

*Note:* * indicates cases where AT&T reached an agreement with the BOC but not with GTE.

**Table A3. Premature Litigation Record**

| State | Arbitration b/w AT&T & GTE Premature Claim Filed by | | Arbitration b/w AT&T & BOCs Premature Claim Filed by | | BOC |
|---|---|---|---|---|---|
| | GTE | AT&T | BOC | AT&T | |
| Alabama | No | No | No | No | BellSouth |
| California | Yes | No | No | No | PacTel |
| Florida | Yes* | No | No | Yes | BellSouth |
| Hawaii | Yes** | No | N/A | N/A | N/A |
| Illinois | Yes | No | No | No | Ameritech |
| Indiana | Yes | No | No | No | Ameritech |
| Iowa | No | No | No | No | US West |
| Kentucky | No | No | Yes* | No | BellSouth |
| Michigan | Yes* | No | No | Yes** | Ameritech |
| Minnesota | Yes* | Yes** | No | No | US West |
| Missouri | Yes* | No | Yes* | No | SBC |
| Nebraska | Yes* | No | No | No | US West |
| New Mexico | No | No | No | No | US West |
| N. Carolina | No | No | No | Yes** | BellSouth |
| Ohio | Yes* | No | No | No | Ameritech |
| Oklahoma | Yes* | No | No | No | SBC |
| Oregon | Yes* | No | Yes* | No | US West |
| | Yes* | No | No | No | Bell Atlantic |
| S. Carolina | No | No | No | No | BellSouth |
| Texas | Yes* | No | No | No | SBC |
| Virginia | Yes* | No | No | No | Bell Atlantic |
| Washington | Yes* | No | No | No | US West |
| Wisconsin | Yes* | No | No | No | Ameritech |
| **Totals** | **17 out of 23** | **1 out of** | **3 out of 22** | **3 out of** | |

*Note:* * dismissed by a federal court as premature; ** withdrawn by the plaintiff.

**Table A4—Lines Provided by ILECs to Competitors for Resale**

| | Resold Residential Lines as % of ILECs' Switched Lines | | | Resold Business and Other Lines as % of ILECs' Switched Lines | | |
|---|---|---|---|---|---|---|
| **State** | **BOC** | **GTE** | company with larger % | **BOC** | **GTE** | Company with larger % |
| Alabama | 1.41 | not reported | n/a | 3.38 | not reported | n/a |
| California | 1.16 | 0.94 | BOC | 1.87 | 0.73 | BOC |
| Florida | 0.90 | 1.14 | GTE | 3.60 | 2.09 | BOC |
| Illinois | 2.01 | 0.00 | BOC | 3.88 | 0.46 | BOC |
| Indiana | 0.34 | 0.06 | BOC | 1.38 | 0.71 | BOC |
| Iowa | 0.14 | not reported | n/a | 2.12 | not reported | n/a |
| Kentucky | 1.46 | 0.12 | BOC | 5.49 | 1.05 | BOC |
| Michigan | 2.31 | 0.00 | BOC | 2.02 | 0.00 | BOC |
| Minnesota | 0.80 | not reported | n/a | 6.71 | not reported | n/a |
| Missouri | 1.11 | not reported | n/a | 2.32 | not reported | n/a |
| Nebraska | 0.62 | not reported | n/a | 1.29 | not reported | n/a |
| New Mexico | 0.00 | not reported | n/a | 0.06 | not reported | n/a |
| North Carolina | 0.67 | 0.01 | BOC | 2.93 | 1.09 | BOC |
| Ohio | 0.21 | 0.00 | BOC | 5.22 | 0.04 | BOC |
| Oklahoma | 2.44 | not reported | n/a | 2.41 | not reported | n/a |
| Oregon | 0.43 | 0.06 | BOC | 0.67 | 0.03 | BOC |
| Pennsylvania | 0.68 | 0.08 | BOC | 2.34 | 0.26 | BOC |
| South Carolina | 3.37 | not reported | n/a | 5.13 | not reported | n/a |
| Texas | 3.26 | 1.08 | BOC | 4.30 | 0.77 | BOC |
| Virginia | 0.17 | 0.00 | BOC | 1.07 | 0.00 | BOC |
| Washington | 0.10 | 0.11 | GTE | 0.44 | 0.08 | BOC |
| Wisconsin | 0.40 | 0.00 | BOC | 4.60 | 0.00 | BOC |
| **# of times BOC has larger %** | | | **12 out of 14** | | | **14 out of 14** |

*Source:* FCC's "Local Competition, August 1999" survey.

**Table A5—Unbundled Loops Provided by ILECs to Competitors**

|  | Unbundled Loops as % of Total Switched Lines | | |
|---|---|---|---|
| **State** | **BOC** | **GTE** | Company with larger % |
| Alabama | 0.10 | not reported | n/a |
| California | 0.26 | 0.13 | BOC |
| Florida | 0.06 | * | BOC |
| Illinois | 0.28 | 0.00 | BOC |
| Indiana | * | 0.00 | BOC |
| Iowa | * | not reported | n/a |
| Kentucky | 0.08 | * | BOC |
| Michigan | 0.88 | 0.00 | BOC |
| Minnesota | 0.09 | not reported | n/a |
| Missouri | 0.08 | not reported | n/a |
| Nebraska | 0.09 | not reported | n/a |
| New Mexico | 0.25 | not reported | n/a |
| North Carolina | 0.08 | * | BOC |
| Ohio | 0.58 | 0.00 | BOC |
| Oklahoma | 0.12 | not reported | n/a |
| Oregon | 0.05 | 0.21 | GTE |
| Pennsylvania | 0.46 | * | BOC |
| South Carolina | * | not reported | n/a |
| Texas | 0.07 | 0.81 | GTE |
| Virginia | * | 0.00 | BOC |
| Washington | * | 0.00 | BOC |
| Wisconsin | 0.32 | 0.10 | BOC |
| **# times BOC (GTE) has larger %** | | | **12 (2) out of 14** |

*Source:* FCC's "Local Competition, August 1999" survey; * amount is positive, but less than 0.05%.

**Table A6—Percentage of ILEC Lines Served by Switching Centers Where Competitors Have Collocation Arrangements**

| | Lines in Switching Centers Where Competitors Have Collocation Arrangements, % of Total Switched Lines | | |
|---|---|---|---|
| **State** | **BOC** | **GTE** | Company with larger % |
| Alabama | 34.7 | not reported | n/a |
| California | 77.3 | 51.8 | BOC |
| Florida | 39.7 | 21.4 | BOC |
| Illinois | 75.7 | 5.5 | BOC |
| Indiana | 46.8 | 20 | BOC |
| Iowa | 47.6 | not reported | n/a |
| Kentucky | 24.6 | 10 | BOC |
| Michigan | 54.4 | 0 | BOC |
| Minnesota | 49.4 | not reported | n/a |
| Missouri | 35 | not reported | n/a |
| Nebraska | 48.4 | not reported | n/a |
| New Mexico | 34.7 | not reported | n/a |
| North Carolina | 48.3 | 19.5 | BOC |
| Ohio | 55.1 | 2 | BOC |
| Oklahoma | 34.4 | not reported | n/a |
| Oregon | 43.1 | 29.8 | BOC |
| Pennsylvania | 50.9 | 7.5 | BOC |
| South Carolina | 18.1 | not reported | n/a |
| Texas | 46.5 | 17.2 | BOC |
| Virginia | 43.8 | 5.2 | BOC |
| Washington | 67.3 | 23.8 | BOC |
| Wisconsin | 86.2 | 2.3 | BOC |
| **# times BOC (GTE) has larger %** | | | **14 out of 14** |

*Source:* FCC's "Local Competition, August 1999" survey.

---

[1] Pub. L. 104-104, 110 Stat. 56, February 8th 1996.

[2] The BOCs had been barred from interLATA services since the 1982 antitrust consent decree, which broke up the vertically integrated AT&T and resulted in the formation of the BOCs. The seven original BOCs were Ameritech, Bell Atlantic, BellSouth, NYNEX, Pacific Telesis, Southwestern Bell and U.S. West. There are now four, following the mergers between Pacific Telesis, Southwestern Bell and Ameritech, and between NYNEX and Bell Atlantic. Pursuant to the antitrust consent decree, Local Access Transport Areas (LATAs) were created and BOCs were confined to only carry calls originating and terminating within the same LATA. There are about 160 LATAs in BOC regions (thus, a state typically contains several LATAs).

[3] See Schwartz (2000) for a detailed discussion of this debate.

[4] GTE was never barred from long-distance services, but had been required to offer such services through a separate subsidiary, under the consent decree arising from its merger with Sprint. (Sprint was subsequently divested by GTE, and the separation requirements imposed on GTE were in litigation at the time of the 1996 Act.) The 1996 Act ended these separation requirements.

[5] There is reason to believe that AT&T's experience is representative of at least the experience of other large IXCs in their role as local entrants negotiating with ILECs. The experience of smaller entrants may be less or more favorable. Smaller entrants could be an easier target for incumbent LECs; but they also are less threatening than the large IXCs, hence ILECs may choose to be more accommodating toward smaller entrants for purpose of "window-dressing" their 271 applications. Comparing the experience of small and large entrants is not attempted here, but would be an interesting extension.

[6] In what follows, by unbundled elements entry mode I mean the case where competitors employ a combination of *owned* and *leased* facilities.

[7] Even an entirely facilities-based entrant needs interconnection between its own, newly established network, and the incumbent's. Absent interconnection, the entrant's subscribers would not be able to communicate with the incumbent's subscribers. Since a network's value to a perspective customer depends critically on the number of persons who can be reached through it, an incumbent could stifle competition from entrants with a small installed base by setting price and non-price terms (e.g., quality) of interconnection, so as to deny an entrant's customers access to the positive networks externalities from communicating with the incumbent's more numerous subscribers. For recent analyses, see Laffont, Rey, and Tirole [1998a, 1998b] and Cremer, Rey, and Tirole [1999]. For a survey of economic literature on network externalities and the relevant bibliography, see Katz and Shapiro [1994], and Besen and Farrell [1994].

[8] Excluded are two states where AT&T did not request interconnection (Alaska and Nevada), two states where AT&T decided to suspend its entry plans (Arkansas and Idaho), and one Arizona—GTE and AT&T agreed that the interconnection agreement reached in California would apply in Arizona as well)

[9] The *loops* are either copper or fiber optic cables that connect the customer premises to the LEC's local offices (end-offices) where computers routing calls are located. *End-Office switching* is the use of the computers (switches) located in the end-offices to route traffic to or from end users to the rest of the telephone network. Entrants might also elect to simply redistribute the ILEC's retail services; in that case discounted wholesale rates for the ILEC's retail services need to be set. The *resale discount rates* are supposed to reflect the retailing costs an ILEC avoids when it sells to a CLEC at wholesale rather than selling to retail customers (costs of billing, marketing, etc.).

[10] In what follows, I report data on the BOCs both individually and as a group. I consider the seven "original" BOCs, that is, before the PacTel-SBC-Ameritech and BellAtlantic-NYNEX merger. However, since GTE is not active in any NYNEX states, no data is reported for NYNEX.

[11] It is reasonable to suppose that GTE too faces some costs of delaying local entry, in the form of regulatory or other sanctions, and that such costs result in an "interior solution" where GTE imposes only a finite delay (e.g., suppose the marginal benefit of delaying entry is constant, while the marginal cost increases with the length of delay, due to the disproportional likelihood or size of regulatory sanctions). This equilibrium delay will be shorter for the BOC since it faces an additional cost of imposing delay, viz., jeopardize its long-distance authority.

[12] In its simplest form, the McNemar test can be used to test equality of the marginal distributions in a two-by-two contingency table (as Table I). The test strategy is based on the observation that, under the null hypothesis, the probability of an observation falling in the top off-diagonal cell equals the probability of an occurrence in the bottom off-diagonal cell. thus, under the null, the frequency in either of the off-diagonal cells is distributed as a binomial $B(n, p)$, where n is the total number of off-diagonal frequencies, and $p=½$.

[13] The Sign Test is based on the observation that, under the null hypothesis, the number of times that an agreement was reached faster with either ILEC is a binomial random variable $B(n,p)$, with $p=½$ (and, in the case at hand, $n=12$).

[14] Here, and in what follows, t tests are carried out as one-tail tests, without assuming equal variances across the two groups of observations. In the case at hand, the statistic had 14 degrees of freedom, and its value was 2.61.

[15] This was possible in all cases except for New Mexico, where information on AT&T's negotiations with both GTE and US West was not available.

[16] Note also that the evidence I gathered about the negotiation process (not reported here for brevity) supports the contention that state commission did not give precedence to negotiation involving the BOCs (of course, within the deadlines for state actions imposed by the Act, which apply to all ILECs).

[17] The purpose of interlocutory orders was to request further studies on unresolved issues, to direct the parties on how to incorporate arbitrated decisions into a complete interconnection agreement, and to set deadlines for submission of a final draft agreement.

[18] I obtained the same results in a censored regression explaining the length of negotiations between AT&T and all ILECs (GTE and the BOCs) through a GTE dummy (which was significant) and a dummy for premature litigation (which was not). Note also that, some of GTE's claims were dismissed as premature by a federal court *after* the state had approved the underlying interconnection agreement (in Florida, Minnesota and Nebraska). This suggests that state commissions sometimes went on with their proceedings without waiting for federal court rulings (although perhaps not as quickly as they would have in the absence of premature litigation).

[19] As for the other three, I was unable to obtain information about their status.

[20] Iowa, Michigan, Minnesota, Pennsylvania and Washington.

[21] The length and outcome of negotiations is arguably the result of AT&T's posture, not only the incumbent LEC's (although, as noted earlier, the evidence does not indicate that AT&T negotiated "harder" with the BOCs).

[22] See Beard, Kaserman and Mayo [1999] for a discussion of the role of resale entry in promoting local competition.

[23] The Hatfield Model was developed by an independent consulting firm, Hatfield Associates Inc., on behalf of MCI and AT&T to provide cost estimates for basic local service on an element-by-element basis. I use the HAI Model Release 5.0a, 1998.

[24] For each company in each state, the Hatfield Model reported the number of residential switched access lines in several categories: residential ($n_1$), business single-line ($n_2$), and business multi-line ($n_3$), along with the actual revenues from providing basic local service in 1996 (R). The FCC's annual "Reference Book of Rates, Price Indices and Expenditures for Telephone Service" reports representative local residential and business rates in 95 urban areas. I used the data for 1997 to compute—net of the federally mandated subscriber line charge and average local taxes—tariffs ratios between: i) business single-line and residential ($r_2$), and ii) business multi-line to the resident, for each company in each state. This was done as follows: when there were one or more urban areas in the FCC's sample for a given company-state combination, I used the corresponding average of $r_1$ and $r_2$; when the FCC had not sampled any company's city in the state, I used the company-wide $r_1$ and $r_2$ averages in the FCC's sample. Thus, I computed the residential average tariff as $R/\kappa$, and the business one as $[R(n_2r_2+n_3r_3)]/[\kappa(n_2+n_3)]$, where $\kappa=n_1+n_2r_2+n_3r_3$.

[25] In this case, $H_0$ is rejected in favor of $H_1$ at the 5% level, and AT&T's requests show no statistically significant bias toward either GTE or the BOCs.

[26] The probability of GTE requesting the higher price/cost ratios 11 or more times is 10.5% under the null hypothesis the BOCs and GTE are equally likely to make the tougher request.

[27] In particular, it has been argued that AT&T might be relatively less keen to reach agreement with the BOCs than with GTE, so as to delay the BOCs' own entry into long-distance. But this hypothesis would predict that AT&T's pricing demands would be tougher to the BOCs than to GTE and that delays in

reaching agreements would be longer with the BOCs (as AT&T would drag its feet). In fact, the data show the opposite.

[28] The link can occur in several ways. BOC resistance to offering prices favorable to entrants could alienate the state commission, which must certify that the BOC has satisfied the "competitive checklist" for section 271. BOC resistance, e.g., through pending appeals of state arbitration decisions, may also lead the FCC independently to conclude that the BOC's local market is not open to competition.

[29] Available from the author upon request.

[30] Following the merger between Southwestern Bell and Pacific Telesis, the information regarding PacTel in California (the only PacTel state in my sample), was reported under Southwestern Bell in FCC Survey.

[31] The FCC has made this information publicly available on its website: http://www.fcc.gov/cbb/stats.

[32] The probability of 12 or more "successes" on 14 tries when the probability of a success is ½ is 0.0065.

[33] The value of test statistic is 93, the lower bound for the 1% critical region is 74.1.

[34] The value of test statistic is 63, the lower bound for the 1% critical region is 52.4.

[35] The FCC reports the percentage of ILEC lines served by switching centers where competitors have collocation arrangements *broken down into "residential" and "other lines"*. For brevity, I report only the aggregate percentage, since in each state where GTE and the BOC provide information, the BOC has the higher percentage for both residential and other lines.

[36] As the data refer to entry as of the end of 1998—barely one month after its merger with Southwestern Bell—I classify SNET as a non-BOC ILEC.

[37] A state-by-state comparison of entry records into BOC and Sprint service areas yields similar results as the comparison between BOC and GTE documented earlier. Sprint's percentage of resold lines is higher than the BOC's in only 2 states out of 15 (Nevada and New Jersey) for residential lines; and only in Tennessee for business resold lines. Sprint provides a higher percentage of UNE loops than the BOC in two states (Nevada and Virginia). Finally, Sprint's collocation record always falls short of the BOC in the same state. Since there is no BOC in Connecticut, no comparison is possible for SNET.

[38] Using FCC data for 1996 (from the *Statistics of Communications Common Carriers* and *Universal Service Support and Telephone Revenue by State*, both available at http//www.fcc.gov), on a state-by-state basis, the correlation between number of toll calls and end-user revenues (intra and interLATA) is 96%.

[39] Information on these variables is available on-line: http://www.census.gov.

[40] In the case of resale entry, one could argue that the explanatory variables capturing subscribers' density and cross-subsidization should not affect entry decisions *directly* (these depend on how resale discounts compare to entrants' marketing costs, regardless of density and the difference between the tariffs ILEC charges to its customers and its actual costs). However, since resale entry is arguably the means to establish customer base with a view to facilities-based competition, these two factors indirectly affect resale entry plans.

[41] Non-BOC ILECs reported no *residential* resold lines in eight states (five GTE: Illinois, Michigan, Ohio, Virginia, and Wisconsin; three Sprint: Indiana, Minnesota, and Washington); no *business* resold lines in five (three GTE: Michigan, Virginia, and Wisconsin; two Sprint: Indiana and Washington).

[42] Estimating the regression of interest separately from the one determining whether the observation belongs to the sample has been referred to as a "two-part model" approach. This approach is justified when the errors in the two equations are independent; otherwise, one should use the Heckman "selection model" approach. Using Monte-Carlo simulations, Manning et al. (1987) show that, even when the data-generating phenomenon would require applying the "selection approach", using the "two-part" model technique leads to estimates that are no worse, and often appreciably better than "selection model" estimates.

[43] The FCC reports the number of loops provided to competitors, but not its break-down into residential and business loops. The information reported thus confounds business and residential entry records; this prevented me from running regressions where the differences between the residential and business segment (number of lines, subscribers density, etc.) are explicitly taken into account.

[44] Similarly to resale regressions, sample selection bias turned out not to affect the estimates. As for the 11 ILECs that reported zero collocation, they were all non-BOC: GTE in Michigan, and Sprint in ten States (Indiana, Kansas, Minnesota, Missouri, New Jersey, Pennsylvania, South Carolina, Tennessee, Texas, and Washington).